\documentclass{PoS}

\title{Top-quark production at hadron colliders}

\ShortTitle{Top-quark production at hadron colliders}

\author{\speaker{Roberto~Bonciani}
         \thanks{Work partly supported by the initiative ``Projets de Physique 
	 Th\'eorique de l'IN2P3''.}\\
        Laboratoire de Physique Subatomique et de Cosmologie,
        Universit\'e Joseph Fourier/CNRS-IN2P3/INPG,
        F-38026 Grenoble, France\\
        E-mail: \email{roberto.bonciani@lpsc.in2p3.fr}}

\author{Andrea~Ferroglia
         \thanks{Work supported by the PSC-CUNY Award N. 64133-00 42 and by the 
	 NSF grant PHY-1068317.}\\
        New York City College of Technology
        300 Jay Street, NY 11201 Brooklyn, US\\
        E-mail: \email{AFerroglia@citytech.cuny.edu}}

\abstract{
 The current theoretical  predictions for the observables related to the top-quark pair 
 and the single-top productions at hadron colliders are briefly reviewed. 
 The theoretical predictions are compared to the experimental measurements carried out at 
 the Tevatron and  the LHC. 
}

\FullConference{XXIst International Europhysics Conference on High Energy Physics\\
                 21-27 July 2011\\
                 Grenoble, Rh\^ones-Alpes France}

\newcommand{\be}{\begin{equation}}
\newcommand{\ee}{\end{equation}}
\newcommand{\bea}{\begin{eqnarray}}
\newcommand{\eea}{\end{eqnarray}}
\newcommand{\bd}{\begin{displaymath}}
\newcommand{\ed}{\end{displaymath}}

\newcommand{\bc}{\begin{center}}
\newcommand{\ec}{\end{center}}

\begin{document}

\section{Introduction}

The top quark is the heaviest elementary particle produced so far at colliders. Due to
its large mass, it is expected to interact strongly with the electroweak symmetry
breaking sector of the Standard Model (SM). Consequently, the top-quark plays a key role
in the investigation of the origin of particle masses, both in the SM and in models of 
``new physics'' (NP).

At hadron colliders, top quarks are produced via two production mechanisms: {\it i)} in
``pairs'', $pp(\bar{p}) \to t\bar{t}$ and {\it ii)} as ``single tops'', together with a
bottom-quark jet, $pp(\bar{p}) \to t\bar{b}$, or a light-quark jet, $pp(\bar{p}) \to
tq(\bar{q})$, or in association with a $W$ boson, $pp(\bar{p}) \to tW$. The pair
production process occurs more than twice as often as the single top production.
Moreover, its experimental signature is cleaner.
For these reasons, the top quark was originally discovered in $t\bar{t}$ events and it
took  14 more years to detect single-top production events at the Tevatron.

The top quark has a very short lifetime: it decays almost exclusively in a $b$ quark and
a $W$ boson in $\sim 5 \cdot 10^{-25}$s. Since the top-quark lifetime is about one order
of magnitude smaller than the hadronization time, the top is the only quark which does
not hadronize. Consequently, the top-quark quantum numbers are accessible to the
experimental measurements. Its spin and the exact nature of its couplings to the $W$
boson can be studied starting from the geometrical distribution of the decay products.
The top-quark mass, which together with the $W$ mass plays an important role in
constraining the Higgs mass via radiative corrections, can be measured with great
accuracy, provided that a satisfactory theoretical definition of this parameter is
employed.

The top-pair and the single-top production processes at hadron colliders can also be
regarded as background for NP processes. In this short review the top quark events are
considered as signal.

In the following we briefly review the status of the measurements of top-quark
observables at the Tevatron and at the LHC, and we summarize the corresponding
theoretical  predictions.

\section{Top-Quark at the Tevatron}

From the top-quark discovery in 1995 \cite{top95} and until the shutdown on September 
30, 2011, the top-quark production and decay have been extensively studied at the 
Fermilab Tevatron.

The channels used at Tevatron for the study of the pair production are three: the
``di-lepton'' channel, in which the two $W$s, which originate from the $t$ and
$\bar{t}$, decay  leptonically; the ``lepton+jets'' channel, in which one of the $W$s
decays leptonically while the other decays hadronically, and the ``all-jets'' channel,
in which both $W$s decay hadronically. In the first case the experimental signature
consists of a pair of high-$p_{T}$ leptons, at least two jets (of which two originate
from bottom quarks), and missing energy due to the neutrinos. In the  ``lepton+jets''
channel, one finds one isolated high-$p_{T}$ lepton, at least four jets (of which  two
are $b$ jets), and missing energy. Finally in the ``all-jets'' channel the  experimental
signature consists of at least six jets, two of which are $b$ jets, and no missing 
energy.
The background processes to $t\bar{t}$ events are: $W+$ jets, di-boson production, 
``all QCD'' events, and  Drell-Yan events. Roughly speaking, the background contains no
$b$ jets. Therefore, a very important  tool to enhance the signal-to-background ratio
is  $b$ tagging.

The single top production events  are characterized by the so-called $t$-channel,
$s$-channel,  and $tW$ associated  production mechanisms. For all of them, events in
which the $W$,  produced by the top, decays leptonically  are usually employed for the
process detection.  The experimental signature for the $t$-channel consists of one
isolated high-$p_{T}$ lepton, with a $b$ jet, a light-quark jet at high  rapidity and
missing energy.  The $s$-channel signature consists in one high-$p_{T}$ lepton, two $b$ 
jets, and missing energy. The associated production cannot be seen at Tevatron.
The background processes for the single top production coincide with the ones for the 
$t\bar{t}$ production and, in addition, the $t\bar{t}$ production itself is part of the 
background for the single top process.

The first top-related observable measured at the Tevatron was the total pair-production
cross section, defined as $\sigma_{t\bar{t}}=(N-N_{bkg})/(\epsilon L)$, where $N$ is the
number of measured events, $N_{bkg}$ is the number of background events, simulated by a
MC event generator, $\epsilon$ is the pair selection efficiency (also simulated with a
MC), and $L$ is the luminosity, measured with data-driven techniques. A recent
measurement of $\sigma_{t\bar{t}}$ by the CDF collaboration gives
$\sigma_{t\bar{t}}^{(CDF)}=7.5 \pm 0.48$~pb \cite{sigmattbarCDF}. The experimental error
corresponds to a relative error of  $\Delta \sigma_{t\bar{t}} / \sigma_{t\bar{t}} \sim
6.5\%$. The measured pair production  cross section is in good agreement with the SM
value (see Section~\ref{thpred}). 
The differential distribution of the top pair production cross section with respect to
the $t \bar{t}$ invariant mass is interesting since  it can  reveal the presence of NP
resonances in the spectrum. The measurement performed by the CDF collaboration is
presented in \cite{inttbarmass}. The distribution is in agreement with the SM prediction
\cite{AhrensInvMass}. The $p_T$ distribution was measured by the D0 collaboration
\cite{ptdistrib} and it also shows a good agreement with the corresponding SM result
\cite{AhrensPTdistr}.

For what concerns the single top production, the current combined Tevatron measurement
of the $(s+t)$-channel cross section is $\sigma_{s+t}=2.76^{+0.58}_{-0.47}$~pb
\cite{SingleCDFD0},  which is in  good agreement with the SM value. This measurement
also allows the determination of $|V_{tb}|$.  However, the CDF collaboration registers a
tension in the ratio of the $s$- and $t$-channel cross sections. The measured ratio is
more than 2$\sigma$ away from the SM prediction \cite{SingleCDF}.
Finally, recently the cross section for the process $p\bar{p} \to t\bar{t}+\gamma$ was
measured by the CDF collaboration \cite{ttbargamma}, finding $\sigma_{t\bar{t}\gamma}=
0.18 \pm 0.08 \, \mbox{pb}$, in agreement with the SM prediction.

The current Tevatron combination of top-quark mass measurements is $m_t=173.2 \pm
0.9$~GeV, with a relative error of only $\sim 0.5\%$ \cite{topmass}. However, given the
fact that the measurement is carried out by comparing data with MC simulations, and
since the top-mass parameter used in the MC is not well defined theoretically, it would
be desirable to have a measurement of $m_t$ related to a well defined Lagrangian
parameter.
Recently, the D0 collaboration evaluated the on-shell and $\overline{\mbox{MS}}$ top
masses by comparing data to the most up-to-dated theoretical predictions for the
production cross section, finding a mean value for the on-shell $m_t$ which is slightly
below the value of 173.2 GeV, but still compatible with it within one standard
deviation  \cite{D0topmasssigma}.
The width of the top quark is also measured at Tevatron. The limit reported by CDF is 
$\Gamma_t < 7.6 \, \mbox{GeV}$ at the 95\% CL, and the value measured by D0 is $\Gamma_t
= 1.99^{+0.69}_{-0.55} \, \mbox{GeV}$ \cite{width}.
The difference between the top and the anti-top masses was measured by the D0
collaboration and it is compatible with zero: $\Delta m_t = 0.8 \pm 1.8 (stat) \pm 0.5
(syst)$GeV \cite{D0massdiff}.

The $W$ helicity fractions, $F_0$, $F_R$ and $F_L$, are measured by  fitting the
one-parameter distribution of the positive charged lepton coming from the $W$ decay.
Using CDF and D0 measurements that simultaneously determine $F_0$ and $F_R$, one finds
$F_0=0.732 \pm 0.081$, $F_R=-0.039 \pm 0.045$ \cite{helW}, in full agreement with the
NNLO  SM predictions \cite{CzarneckiNNLO}.
Also the spin correlations are in agreement with the SM predictions
\cite{BernreutherSi}. D0 measures a $t\bar{t}$ spin correlation strength, using as spin
quantization axis the direction of the beam, of $C=0.1^{+0.45}_{-0.45}$, while CDF finds
$C = 0.72 \pm 0.64(stat) \pm 0.26(syst)$ \cite{spincorr}. However, the error bands are
too big for any  claim.

Tevatron experiments are also searching for NP in top-quark pair and single-top
production processes. This activity includes searches for new resonances, as for
instance a $Z'$ or  a $W'$, for possible anomalous couplings of the top quark to the $W$
that can reveal a discrepancy with respect to the $V-A$ structure of the SM, the search
for a fourth generation (with decays in SM particles) or for non-SM decays of the top
quark, as for instance $t \to H^+b \to q \bar{q}' b (\tau \nu b)$. 
However, so far, no evidence of NP was found, and the good agreement with the SM
predictions is used to set constraints on the NP parameters, such as the masses and the
couplings of the conjectured NP particles. 

The only observable which exhibits a sizable discrepancy with respect to the 
corresponding SM prediction is the top-pair forward-backward asymmetry, $A_{FB}$. This
observable is defined as $A_{FB}^{(i)}=
(N_t(y_t>0)-N_t(y_t<0))/(N_t(y_t>0)+N_t(y_t<0))$, where $N_t(y_t>0)$ ($N_t(y_t<0)$) is
the number of top quarks with positive (negative) rapidity, and $i$ labels the frame of
reference in which the measurement of the rapidity is carried out. The asymmetry is
measured  either in the laboratory frame or in the $t \bar{t}$ rest frame. Assuming CP
invariance (the SM CP violation is irrelevant here), the forward-backward asymmetry
coincides with the charge asymmetry
$A_C^{(i)}=(N_t(y_t>0)-N_{\bar{t}}(y_{\bar{t}}>0))/(N_t(y_t>0)+N_{\bar{t}}(y_{\bar{t}}>0))$,
since we have $N_{\bar{t}}(y_{\bar{t}}>0)=N_t(y_t<0)$.
The measurement of the forward backward asymmetry in the $t \bar{t}$ rest frame can be
carried out by employing the fact that the difference between the top and anti-top 
rapidities $\Delta y = y_t - y_{\bar{t}}$ is invariant with respects to boosts along
the  beam axis, and that  it is by definition equal to twice the top-quark rapidity in
the $t \bar{t}$ frame: $\Delta y = 2 y_t^{t \bar{t}}$. Consequently, one find that
$A_{FB}^{(t\bar{t})}= (N_t(\Delta y>0)-N_t(\Delta y<0))/(N_t(\Delta y>0)+N_t(\Delta
y<0))$.
The measurement of the asymmetry is obtained by employing  ``lepton+jets'' events. The
charge of the observed lepton allows one  to determine whether the  parent parton is a
top or an anti-top, and therefore to know also the nature of the other  top in the pair.
The latter gives origin to the $W$ boson which decays hadronically.   The momentum of
the top quarks which decays hadronically can be reconstructed, since  all of the jets
originating from the decay are detected.  CDF collaboration also uses ``di-leptonic''
events. In this case the analysis is complicated by the missing energy due to the
neutrinos. 
CDF and D0 collaborations find comparable forward-backward asymmetry values, 
$A_{FB}^{(t\bar{t}),CDF} = 0.201 \pm 0.067$ and  $A_{FB}^{(t\bar{t}),D0} = 0.196 \pm
0.065$ \cite{AFB}, which are more than $2\sigma$ larger than the SM theoretical
predictions at the NLO \cite{AFBtheory}.  The asymmetry value  in the laboratory frame
measured by CDF is $A_{FB}^{(lab),CDF} = 0.150 \pm 0.055$ \cite{AFBmttbar}. Moreover,
CDF registers a further discrepancy with the SM value in the dependence of 
$A_{FB}^{(t\bar{t})}$ with respect both to the $t\bar{t}$ invariant mass and rapidity
difference. While for $m_{t\bar{t}} < 450 \, \mbox{GeV}$ the CDF result is compatible
with  the SM prediction within one standard deviation, for $m_{t\bar{t}} > 450 \,
\mbox{GeV}$ the  measured value, 
$A_{FB}^{(t\bar{t})}(m_{t\bar{t}} > 450 \, \mbox{GeV}) = 0.475 \pm 0.114$ is $3.4
\sigma$ bigger \cite{AFBmttbar} than the SM prediction. The same happens for the
dependence on the rapidity difference.  At small $\Delta y$, we find 
$A_{FB}^{(t\bar{t})}(|\Delta y| \leq 1.0) = 0.026 \pm 0.104 \pm 0.055$ compatible with
the SM prediction. At large rapidity difference, CDF finds $A_{FB}^{(t\bar{t})}(|\Delta
y| \geq 1.0) = 0.611 \pm 0.210 \pm 0.141$ \cite{AFBdeltay},  which is much larger than
the SM prediction  $A_{FB}^{(t\bar{t})}(|\Delta y| \geq 1.0) = 0.123 \pm 0.018$
\cite{mcfm}. However, neither  the increase of the asymmetry for large values of the
pair invariant mass, nor the increase of the asymmetry for  large values of  $|\Delta
y|$ is at the moment confirmed by $D0$ \cite{AFB}. 
For a dedicated review on possible NP explanations of the $A_{FB}$ discrepancy we
refer  the reader to \cite{Westhoff}.

\section{Top-Quark at the LHC}

Since the end of 2010, also CMS and ATLAS collaborations at the LHC are producing
accurate measurements of the top properties. The most recent values are based on  $\sim
1 \,\mbox{fb}^{-1}$ of data, recorded between the end of 2010 and Summer 2011.

The $t\bar{t}$ production cross section was measured by both collaborations: the
measured values are $\sigma_{t \bar{t}}^{(CMS)} = 158 \pm 19 \, \mbox{pb}$ and
$\sigma_{t \bar{t}}^{(ATLAS)} = 176 \pm 5(stat) ^{+13}_{-10} (syst) \pm 7(lum) \,
\mbox{pb}$ \cite{CSLHC}.  Therefore, after only few months of data taking, the relative
error on this observable is already quite small ($\sim \! 10-\!15\%$); furthermore the
statistical uncertainty is already smaller than the systematic one.
The t-channel single top production cross section is measured with a larger relative
error of $\sim 30\%$: $\sigma_{t}^{(CMS)} = 83.6 \pm 29.8(stat + syst) \pm 3.3 (lum) \,
\mbox{pb}$   and $\sigma_{t}^{(ATLAS)} = 90{~}^{+32}_{-22} \, \mbox{pb}$ \cite{CSstLHC}.

The top-quark mass was measured by the ATLAS collaboration using a template method
(which suffers of the same problems already pointed out in the previous section): the
value obtained is $m_t^{(ATLAS)} = 175.9 \pm 0.9(stat.) \pm 2.7 (syst.) \, \mbox{GeV}$
\cite{ATLAS-CONF-2011-120}. The CMS collaboration repeated the analysis done by D0,
using the theoretical cross section and measuring the on-shell and
$\overline{\mbox{MS}}$ top masses \cite{CMS-PAS-TOP-11-008}, finding comparable results.
The difference between the top and anti-top masses was measured by CMS, which obtained  a value
compatible with zero: $\Delta m_t = -1.2 \pm 1.2(stat.) \pm 0.5 (syst.) \, \mbox{GeV}$
\cite{CMS-PAS-TOP-11-019}.

$W$-helicity fractions and spin correlations, measured by ATLAS, are also compatible
with their SM value: $F_0 = 0.75 \pm 0.08$, $F_L = 0.25 \pm 0.08$ (setting $F_R=0$)
\cite{ATLAS-CONF-2011-122},  and $\kappa = 0.34^{+0.15}_{-0.11}$ in the helicity base 
\cite{ATLAS-CONF-2011-117}.

Finally, although LHC is a machine with a symmetric initial state  and therefore the
$A_{FB}$ measured at the Tevatron cannot be seen, one can define and measure a charge
asymmetry by exploiting the fact that the rapidity distributions of top and anti-top
quarks are different. The antitops tend to be produced at small rapidity, while the
distribution of the tops is broader. The relevant observable is defined as $A_C =
(N(\Delta |y|>0)-N(\Delta |y|<0))/( N(\Delta |y|>0)+N(\Delta |y|<0))$, with $\Delta
|y|=|y_t|-|y_{\bar{t}}|$ and $y_t$ is the top rapidity; the measured values for this
quantity are $A_C^{(CMS)} = -0.013 \pm 0.026 (stat)^{+0.026}_{-0.021}(syst)$
\cite{CMS-PAS-TOP-10-010} and  $A_C^{(ATLAS)} = -0.024 \pm 0.016 (stat) \pm 0.023(syst)$
\cite{ATLAS-CONF-2011-106}. Due to  the large errors, both values are compatible with
the SM prediction, which is $\sim +1\%$.

The already remarkable accuracy of  the LHC measurements is going to improve in the next
years. For example, in the high-luminosity and  high-energy phase,  the $t\bar{t}$
production cross section is expected to be  measured with an accuracy of 5\%, while the 
single-top $t$-channel cross section will be measured with an accuracy of 10\%.  These
very precise experimental measurements must be matched by equally accurate theoretical
predictions.

\section{Theoretical Predictions \label{thpred}}

The production of a top-antitop pairs is dominated by the strong interaction. The
inclusive  production cross section can be written using the QCD Factorization Theorem
as  
\be
\sigma^{t \bar{t}}_{h_1,h_2}(s_{\mbox{{\tiny had}}},m_t^2) = \sum_{ij} \int_{4
m_t^2}^{s_{\mbox{{\tiny had}}}} d \hat{s}\,  L_{ij}\left(\hat{s},
s_{\mbox{{\tiny had}}}, \mu_f^2 \right) \, 
\hat{\sigma}_{ij}(\hat{s},m_t^2,\mu_f^2,\mu_r^2)  \, , 
\label{CS}
\ee  
where the hard scattering of the partons $i$ and $j$ ($i,j \in \{q , \bar{q}, g\}$) at a
partonic center of mass energy $\hat{s}$ is described by the partonic cross section,
$\hat{\sigma}_{ij}$, which can be calculated in perturbative QCD. The process
independent partonic luminosity, $L_{ij}$, describes the probability of finding, in the
hadrons $h_1$ and $h_2$ (where $h_1, h_2 = p, \bar{p}$ at the Tevatron, while $h_1, h_2
= p, p$ at the LHC), an initial state involving partons $i$ and $j$ with the given
partonic energy $\hat{s}$. The integration extends up to the collider hadronic c. m.
energy $s_{\mbox{{\tiny had}}}$. $\mu_f$ and  $\mu_r$ indicate the renormalization and
factorization scales.

At  leading order  (LO) in perturbation theory, there are two partonic channels
contributing to the  $t\bar{t}$ production cross section: the quark-antiquark channel $q
\bar{q} \to t \bar{t}$ and the gluon fusion channel $gg \to t \bar{t}$. Because of the
interplay between parton luminosity and partonic cross sections, the quark-antiquark
channel dominates the pair production cross section at the Tevatron, while at the LHC
the inclusive cross section is largely dominated by gluon fusion events.

In single-top production, there are three LO partonic channels: {\em i)} $q(\bar{q})b
\to q'(\bar{q}')t$, in which a $W$ boson is exchanged in the $t$ channel, {\em ii)}
$q\bar{q}' \to t\bar{b}$, in which the $W$ boson is exchanged in the $s$ channel, and
{\em iii)} the ``associated $tW$ production''  $gb \to tW$. The $t$-channel process
dominates the single top production both at the Tevatron and at the LHC. The $s$-channel
production was detected at the Tevatron and it plays no role at the LHC. The associated
production, instead, cannot be revealed at the Tevatron while is the second most
important single-top production mechanism at the LHC.

\subsection{NLO Calculations}

The LO predictions for the $t\bar{t}$ or single-top production cross sections are
affected by a huge dependence on the renormalization/factorization scales, and they
cannot be regarded as reliable predictions.  More accurate predictions can be  obtained
by taking into account the NLO corrections, which consist of two parts: the virtual
corrections,  originating from the interference of the one-loop diagrams with the
tree-level ones, and the real radiation corrections, originating from the interference
of the $2\to3$ amplitudes.
For totally inclusive quantities, as for instance the total cross section, one has to
integrate the final-state particles over the complete phase space. This is the approach
used for instance in \cite{CM}. In so doing, the IR divergences of the virtual part
cancel exactly (analytically) against the divergences which originate from the
integration of the additional parton in particular regions of the phase space. 
However, in order to compare directly the theoretical predictions  with the experimental
measurements, one needs to impose cuts and to take into account the geometrical
acceptance of the detectors. Consequently, for more exclusive observables a subtraction
scheme is needed to ``regularize'' the IR collinear and soft divergences, coming from
the integration over the phase space.  The basic idea is the following. The NLO cross
section for the production of $n$ partons in the final state is given by the sum of the
virtual part, integrated over the $n$-particle phase space, and the real part,
integrated over the $n+1$-particle phase space: $\sigma_{NLO} = \int_n \sigma_V +
\int_{n+1} \sigma_R$. The UV divergences are removed by the renormalization procedure.
However, the virtual part exhibits IR divergences, that appear as poles in $\epsilon =
(4-D)/2$, where $D$ is the dimension of the space-time. The same  divergences, with
opposite sign, arise after the integration of the real radiation over the phase
space. 
In order to locally regularize the IR divergences, one adds (and subtracts) a term $\sigma_{S}$
that reproduces the matrix element behavior in all singular limits and, at the same
time, is sufficiently simple to be integrated analytically: $\sigma_{NLO} = \int_n (
\sigma_V - \int_1 \sigma_{S} ) + \int_{n+1} ( \sigma_R + \sigma_{S} )$. The integration
$\int_1 \sigma_{S}$ reproduces the poles of the virtual part, while the
integration of the $n+1$ final state partons is now finite and it can be carried out
numerically in 4 dimensions.
At  NLO, the subtraction terms are completely known and several subtraction
formalisms  are available in the literature \cite{Subtrac}.

\subsubsection{Stable Tops}

Let us first consider the $t$ and $\bar{t}$ in the final state as stable on-shell
particles. The NLO QCD corrections to the total $t\bar{t}$ cross section, summed over
the final spins and colors, were calculated by many groups \cite{Nason:1987xz}. They
enhance the cross section by almost 25\% at the Tevatron and by 50\% at the LHC. The
residual renormalization/factorization scale dependence, plus parton distribution
functions uncertainties, is about 15-20\%. The EW corrections are also known 
\cite{QCDEW}, but their contribution (+1\% at Tevatron and -0.5\% at the LHC) is
negligible in comparison to the QCD theoretical error.

The NLO QCD corrections to the $t$-channel single-top production are moderate. They
enhance  the cross section by 9\% at the Tevatron and by 5\% at the LHC
\cite{topNLOQCDtchannel,Harris}. The NLO EW corrections decrease the cross section by 
1\% both at the Tevatron and the LHC \cite{topNLOEWtchannel}. The NLO QCD corrections to
the  $s$-channel cross section are large, resulting in an enhancement of 47\% at the
Tevatron  and 44\% at the LHC \cite{Harris,topNLOQCDschannel}. Finally, the NLO QCD
corrections  to the associated $tW$ production enhance the cross section by 10\% at the
LHC \cite{topNLOQCDtW}.

For what concerns processes with additional particles in the final state, the NLO
corrections to $t\bar{t}+j$ were calculated in \cite{ttj} (the calculated cross section
at the Tevatron, $\sigma_{t\bar{t}j} = 1.79^{+0.16}_{-0.31}~$pb is in good agreement
with the CDF measurement \cite{CDFttj}) and those to $t\bar{t}+2j$ in \cite{ttjj}. Moreover, the
$t\bar{t}b\bar{b}$ production was considered in \cite{ttbb}. The NLO corrections to the
production of a top pair in association with a photon were calculated in
\cite{Melnikov:2011ta}.

NLO corrections to many processes concerning both $t\bar{t}$-pair and single-top
productions  are matched with parton showers in the publicly available codes MC@NLO
\cite{MCatNLO} and POWHEG \cite{POWHEG}.

\subsubsection{Resummation}

The QCD corrections to processes that involve at least two large energy scales (the
partonic energy $\sqrt{\hat{s}}$ and the top mass $m_t$ are both much larger than
$\Lambda_{QCD}$) are characterized by a logarithmic behavior in the vicinity of the
boundary of the phase space: $\sigma \sim \sum_{m,n} C_{mn} \alpha_S^m \log^n(\rho)$,
where $\rho$ is the kinematic variable that ``measures'' the distance from the exclusive
boundary. When $\rho \ll 1$, even if the transverse momentum is such that $\alpha_S(Q^2)
\ll 1$ and perturbative QCD can be employed, one can find that  $\alpha_S^m \log^n(\rho)
\sim {\mathcal O}(1)$. The logarithmically enhanced terms spoil the convergence of the
fixed-order expansion, that has to be recovered by performing a  systematic resummation
of these terms to all orders in perturbation theory \cite{Sterman:1986aj}. For the
$t\bar{t}$ pair production process, the resummation of the leading logarithmic terms
(LL) was carried out in  \cite{LL}, and the next-to-leading logarithmic (NLL) in
\cite{NLL}, both for the cross section at the production threshold and for the invariant
mass distribution. Recently,  the NNLL resummation was carried out  for the cross
section at threshold, the top-pair invariant mass distribution,  the top-quark
transverse momentum distribution,  and top-quark rapidity distribution
\cite{APPROX,NNLL,AhrensInvMass,AhrensPTdistr}. Different approaches employing either 
Mellin space resummation or  momentum space resummation  techniques based on Soft
Collinear Effective Theory were employed in these works.
For the single top production, the NLL terms were calculated in \cite{kidonakisetc} and
the NNLL in \cite{kidonakis}. A comprehensive review of the recent results obtained with
resummation  techniques can be found in \cite{reviewNNLL}.

\subsubsection{Factorisable Corrections}

In the papers reviewed in the last two sections, the top quarks are treated as stable
particles. However, only hadrons and leptons are detected experimentally and it is on
these particles that the experimental cuts are imposed. Therefore, it is highly
desirable to consider the actual final state  in the theoretical analysis. This is  very
difficult from the point of view of the calculation, since one needs to deal with
Feynman diagrams with many external legs. A first step towards this goal consists in
working within  the ``narrow-width approximation'': since the top-quark behaves as a
narrow resonance, {\it i.e.} $\Gamma_t/m_t \ll 1$, one can formally take the limit
$\Gamma_t/m_t \to 0$ of the complete cross section. 
The limit $\Gamma_t/m_t \to 0$ decouples the  top-quark production process  from the
top-quark decay. 
This approach allows one to compute realistic distributions and to keep trace of the
spin of the tops. It was applied both to the top-pair and the single-top production
processes.

Two groups performed detailed studies of the top-quark pair production within the narrow
width approximation approach \cite{BernreutherSi,Bernreutheretc,Melnikov:2009dn}. For
single-top production, the same formalism was used to study the $t$-channel cross
section at the LHC \cite{Schwien}. Recently, also the off-shell effects both for $t$-
and $s$-channel cross sections at the Tevatron and at the LHC were computed 
\cite{ttbboffshell}.

\subsubsection{Non-Factorisable Corrections for Pair Production}

In 2010, two groups  calculated the complete set of corrections to $pp(\bar{p}) \to
W^+W^-b\bar{b}$, including also the non-factorisable corrections \cite{ppWWbb}. 
The calculation is extremely challenging and it involves  almost 1500 Feynman diagrams with
six external legs. As a by-product of the calculation, the authors could prove that for 
inclusive quantities the non-factorisable corrections are indeed of   ${\mathcal
O}(\Gamma_t/m_t) \sim 1\%$. With these results many exclusive observables,  with
realistic experimental cuts, can be evaluated.

\subsection{Towards a NNLO Analysis of the Top-Pair Production in Perturbative QCD}

The foreseen accuracy with which the LHC will be able to measure some of the top-pair
production observables is such that in several cases the calculation of the NNLO
corrections is required.
While in the single-top production the NLO theoretical predictions (supplemented by the
soft gluon resummation) already match the expected experimental accuracy, this is not
the case for the $t\bar{t}$ production process. In the latter case the inclusion of the 
NNLO corrections in the analysis is needed.
Due to the complexity of the calculations, to date the complete set of NNLO QCD
corrections is not yet available. However, many partial results are known, and the full
calculation of the NNLO correction appears to be within reach.

The most accurate theoretical predictions currently employed for comparison with the
experimental measurements include the ``approximate NNLO'' corrections (NLO plus some or
all of the following ingredients: scale dependence at NNLO, Coulomb terms up to two
loops, logarithmic terms obtained by re-expanding NNLL formulas), both for $t\bar{t}$ 
\cite{APPROX} and for single top \cite{KidoSingletop} productions.

Many parts of the full top-pair NNLO matrix element are known. In \cite{smallmass} the
matrix elements for the $q\bar{q}$ and $gg$ channels were computed in the $s \gg m_t^2$
limit. In \cite{Czakon:2008zk}, matrix elements in the $q\bar{q}$ channel were computed
numerically and by retaining the full  dependence on the top-quark mass. In
\cite{Ferroglia:2009ep}, all of the IR two-loop poles, both in the $q\bar{q}$ and $gg$
channels, were evaluated analytically. In \cite{Korner:2005rg}, the virtual one-loop
times one-loop matrix elements were calculated. Finally, in \cite{quarkloops} the
two-loop fermionic and leading color corrections to the $q\bar{q}$ channel and the
two-loop leading color corrections to the $gg$ channel were computed analytically by
employing the Laporta algorithm \cite{Laportaalgorithm} (as implemented in the {\tt 
C++} code Reduze \cite{Studerus:2009ye}), and the differential equation method
\cite{DiffEq}. While the most complicated four-point master integrals  were evaluated
especially for these projects (see \cite{Andreas}), part of the needed master integrals
were already available in the literature \cite{MIs}.

The computation of exclusive observables at the NNLO requires a subtraction scheme for
the real corrections in presence of massive partons. A complete NNLO subtraction scheme
applicable to the top-pair production is not yet available. However, many intermediate
results were recently obtained. The approach employed follows closely the one adopted
at the NLO. In order to locally regularize the IR divergences that originate from the
integration over the phase space, one adds and subtracts terms that reproduce the
behavior of the matrix element in all the singular limits. 
At NNLO the structure of the singularities due to unresolved partons in the final state
is more involved with respect to the NLO \cite{singNNLOtree,singNNLO1L}. One encounters 
double unresolved singularities and overlapping singularities. Furthermore, when 
integrating the subtraction terms, one needs to evaluate complicated two-loop integrals.
Subtraction terms, together with their integrated counterpart, were published so far
in different frameworks \cite{RealAnalyt}. 
Numerical and semi-analytical techniques, based on sector decomposition, were also 
applied to this problem \cite{RealNum}.

\end{document}